\documentclass[]{fairmeta}

\usepackage{amsmath}
\usepackage{alltt}
\usepackage{enumitem}
\usepackage{pifont}
\usepackage{listings}
\usepackage{float}
\newcommand{\cmark}{\ding{51}}
\newcommand{\xmark}{\ding{55}}
\lstdefinestyle{apcode}{language=Python, basicstyle=\ttfamily\footnotesize\color{metafg}, keywordstyle=\color{metablue}, stringstyle=\color{black!60}, emph={auto_parallel, init_device_mesh}, emphstyle=\color{metablue}\bfseries, backgroundcolor=\color{metabg}, xleftmargin=4pt, framexleftmargin=4pt, aboveskip=4pt, belowskip=4pt, columns=fullflexible, keepspaces=true, showstringspaces=false}

\title{AutoParallel: Composable, PyTorch-Native Automatic Parallelism for Production Training}

\author[1,2,*]{Kaijian Wang}
\author[1]{Sanket Purandare}
\author[1]{Aditya Venkataraman}
\author[1]{Will Constable}
\author[1]{Ivan Kobzarev}
\author[1]{Elias Ellison}
\author[1]{Simon Fan}
\author[1]{Edward Z. Yang}
\author[1]{Davide Italiano}
\author[1]{Patrick Labatut}
\author[2]{Yuke Wang}
\author[1,\dagger]{Francisco Massa}

\affiliation[1]{Meta Superintelligence Labs}
\affiliation[2]{Rice University}
\contribution[\dagger]{Project lead}
\contribution[*]{Work done during an internship at Meta}
\date{October 30, 2026}
\metadata[Code]{\url{https://github.com/meta-pytorch/autoparallel}}
\hypersetup{pdftitle={AutoParallel: Composable, PyTorch-Native Automatic Parallelism for Production Training}, pdfauthor={Kaijian Wang, Sanket Purandare, Aditya Venkataraman, Will Constable, Ivan Kobzarev, Elias Ellison, Simon Fan, Edward Z. Yang, Davide Italiano, Patrick Labatut, Yuke Wang, Francisco Massa}}

\abstract{
Training large models requires multiple interacting forms of parallelism. As models and clusters scale, finding and maintaining efficient parallelization plans becomes essential for training performance. Diverse model architectures, workloads, and hardware topologies demand flexible strategies, yet production PyTorch systems typically rely on hand-written recipes that are costly to adapt and tend to reuse familiar patterns, which may be suboptimal for a given workload. We present \textbf{AutoParallel}, a PyTorch-native planner for distributed model training. AutoParallel searches parallelization choices jointly across the training computation, allowing established techniques to be combined and unconventional strategies to emerge. It expresses its decisions using native PyTorch abstractions and produces ordinary PyTorch programs, avoiding a separate execution runtime while composing directly with the PyTorch compilation and training stack. AutoParallel also composes automatic decisions with model-specific distributed code, allowing dynamic or specialized regions to remain under user control. Plans are inspectable and replayable artifacts. Across dense, hybrid, and mixture-of-experts models on meshes of up to four dimensions and 128 NVIDIA H100 GPUs, AutoParallel matches or exceeds expert-written TorchTitan recipes on most configurations, with throughput gains of up to 18.3\%, and discovers plans that uniform expert recipes do not express.
}

\begin{document}

\maketitle

\FloatBarrier

\section{Introduction}\label{sec:introduction}

Model training has grown at an extraordinary pace. The Hugging Face Hub alone hosts over three million models~\citep{tazi2025ultrascale}; inside organizations, teams routinely train and fine-tune models for tasks spanning natural language, vision, multimodal generation, and more. Widely adopted frameworks such as PyTorch~\citep{paszke2019pytorch}, combined with open model architectures and community libraries, have lowered the barrier to training models.

Achieving \emph{efficient} training, however, is far from straightforward. Scaling model training beyond a single accelerator requires carefully combining multiple forms of parallelism, including Data Parallelism~\citep{rajbhandari2020zero, zhao2023pytorch}, Tensor Parallelism~\citep{shoeybi2019megatron, korthikanti2023reducing}, Context Parallelism~\citep{liu2023ring}, Pipeline Parallelism~\citep{huang2019gpipe, narayanan2019pipedream}, and Expert Parallelism~\citep{fedus2022switch}, together with system-level optimizations including activation recomputation~\citep{chen2016training, korthikanti2023reducing}, mixed precision~\citep{micikevicius2018mixed}, and deep learning compilers~\citep{ansel2024pytorch}. Moreover, identifying an effective training configuration is highly context-specific: it depends jointly on model architecture, workload shapes, and hardware topology~\citep{narayanan2021efficient}.

Hand-tuning a distributed training recipe requires dedicated infrastructure expertise and iterative experimentation, and this cost is amortized only when a single configuration is reused across a long production run. In practice, this makes hand-optimization economical for a narrow regime: a small number of flagship models, each trained by a dedicated team over weeks or months. It does not scale to the broader landscape of model training, where researchers iterate on new architectures faster than recipes can be tuned, product teams train many models for different tasks without a dedicated infrastructure effort for each, and open-source practitioners need efficient training out of the box. A recent benchmarking study by Hugging Face~\citep{tazi2025ultrascale} illustrates the scale of this challenge: of 3{,}306 parallelism configurations for LLaMA models on up to 512 H100 GPUs, nearly half failed with out-of-memory errors, and the sweep cost approximately \$469K while covering only a narrow slice of the configuration space.

A substantial body of prior work~\citep{jia2019beyond, shazeer2018mesh, zheng2022alpa, xu2021gspmd, miao2023galvatron, jia2022whale, unger2022unity, lin2024nnscaler, chen2024uniap} has demonstrated that automatic parallelization is feasible. Despite these advances, automatic parallelism has not replaced manual recipe development in practice. The key barriers are threefold: (1)~many prior systems rely on non-PyTorch runtimes or specialized compiler stacks, placing them outside the frameworks most practitioners actually use; (2)~they often produce opaque solutions that are difficult to inspect, modify, or debug, offering no guarantee of matching the quality of hand-tuned production recipes; and (3)~they compose poorly with the optimizations that production training requires, including compilation, activation checkpointing, and mixed precision. Moreover, modern training recipes increasingly involve dynamic parallelism patterns, such as context parallelism and data-dependent expert routing in mixture-of-experts layers. These patterns require specialized communication primitives beyond static operator placement, and existing systems that operate on static computation graphs cannot readily support them. As a result, although PyTorch is widely used for both research and production training~\citep{paszke2019pytorch, ansel2024pytorch}, PyTorch practitioners still lack an automatic path from model definition to efficient distributed training (\S\ref{sec:motivation}, \S\ref{sec:related-work}).

\begin{figure}[htbp]
\centering
\includegraphics[width=\textwidth]{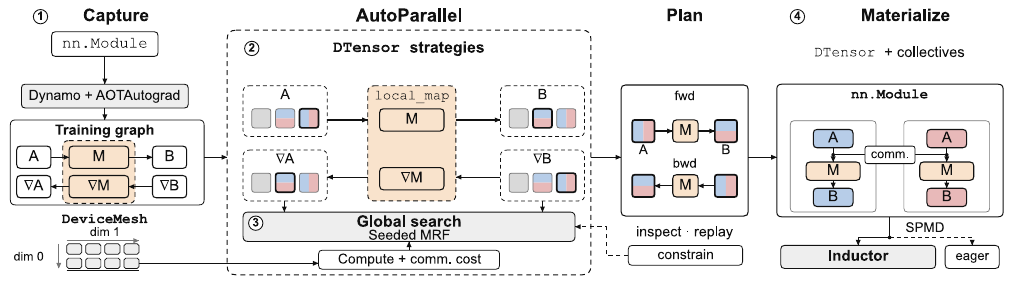}%
\caption{AutoParallel system overview. Colored boxes schematically illustrate tensor sharding, and bold outlines mark the strategy selected for each operator. The orange \texttt{local\_map} region keeps its user-declared placements, and user constraints feed back into the search.}
\label{fig:overview}
\end{figure}

In this paper, we present \textbf{AutoParallel}, a PyTorch-native single-program, multiple-data (SPMD) planner for distributed model training that closes this gap. As Figure~\ref{fig:overview} shows, we add a planning layer between graph capture and plan materialization in the PyTorch stack: it enumerates valid DTensor placements for every operator in the joint forward-backward graph, selects one per operator under a hardware-aware cost model, and emits an ordinary PyTorch module that the existing compiler stack executes without a separate runtime (\S\ref{sec:production-integration}). For dynamic parallelism patterns such as context parallelism and mixture-of-experts, we provide composition boundaries that let manually parallelized regions interoperate with auto-planned computation (\S\ref{sec:composition}). AutoParallel is open source\footnote{\url{https://github.com/meta-pytorch/autoparallel}}~\citep{autoparallelrepo}, has been presented at the PyTorch Conference~\citep{massa2025ptc, graphtrainer}, and is integrated into TorchTitan's GraphTrainer for production training~\citep{torchtitanautoparallel}.

We evaluate AutoParallel on dense, hybrid, mixture-of-experts, and long-context workloads across multi-dimensional meshes and up to 128 NVIDIA H100 GPUs, comparing against expert-written TorchTitan~\citep{liang2024torchtitan} recipes under matched meshes and runtime optimizations (\S\ref{sec:evaluation}). AutoParallel matches or exceeds the expert recipes on most configurations without per-model tuning, and when the workload changes, replanning recovers the performance that a stale plan leaves behind. This paper makes the following contributions:
\begin{itemize}
\item \textbf{Graph-wide placement planning.} We formulate SPMD parallelization as a single optimization over the joint forward-backward graph under a hardware-aware cost model. We introduce a scalable approximate solver that produces near-optimal plans on instances small enough to verify against an exact solver and scales to meshes where exact solving is infeasible.
\item \textbf{Native recipe generation.} We materialize decisions as an ordinary PyTorch program with no separate runtime and integrate it with TorchTitan's GraphTrainer~\citep{graphtrainer, torchtitanautoparallel} for full-graph Inductor compilation with communication-computation overlap and collective bucketing.
\item \textbf{Practical control and composition.} We make plans inspectable and replayable, allow users to constrain placements, and compose them with user-controlled dynamic parallelism regions (context parallelism, mixture-of-experts) via explicit placement contracts (\S\ref{sec:system-overview}, \S\ref{sec:production-integration}).
\item \textbf{Production evaluation.} We report end-to-end performance against expert recipes, replanning across sequence lengths, planner scalability, and cost-model fidelity (\S\ref{sec:evaluation} and Appendix~B).
\end{itemize}

\FloatBarrier

\section{\texorpdfstring{Background and Challenges}{Background and Challenges}}\label{sec:background}

\subsection{Parallelism as Per-Operator Placement}\label{sec:placement-view}

A training computation can be represented as a dataflow graph of tensor operators spanning both the forward and backward passes. A \emph{device mesh} is a multi-dimensional array of accelerators, where each axis corresponds to a parallelism dimension with its own communication bandwidth. AutoParallel targets single-program, multiple-data (SPMD) execution~\citep{xu2021gspmd, lepikhin2020gshard}: every device in the mesh runs the same program on its local shard of each tensor, and devices synchronize through collective communication. A \emph{placement} assigns each tensor one of three types along each mesh axis: \textbf{Replicate} (the full tensor is present on every device along that axis), \textbf{Shard(\emph{d})} (the tensor is partitioned along dimension~$d$), or \textbf{Partial} (each device holds a partial result awaiting reduction). On a $D$-dimensional mesh, a tensor's placement is a $D$-tuple drawn from these types. Under this view, named parallelism strategies such as FSDP, tensor parallelism, and sequence parallelism are not first-class concepts but emergent consequences of per-operator placement decisions within a single SPMD program; \S\ref{sec:strategy-space} develops this correspondence with concrete examples. Data-parallel axes always shard the input batch; on a DDP axis the weights are replicated, and on a DPS axis they are sharded as in FSDP. HSDP combines a DDP axis with a DPS axis.

\subsection{PyTorch Compilation and Distribution Stack}\label{sec:pytorch-stack}

AutoParallel builds on three layers of existing PyTorch infrastructure rather than introducing a separate intermediate representation or runtime. \textbf{TorchDynamo and AOTAutograd}~\citep{ansel2024pytorch} capture the model's forward computation and automatically derive the backward pass, producing a single joint computation graph at the ATen operator level. \textbf{DTensor and DeviceMesh}~\citep{liang2024torchtitan} provide the placement vocabulary and redistribution semantics: each distributed tensor carries its mesh and placement metadata, and transitions between placements are realized through collectives. SimpleFSDP~\citep{zhang2024simplefsdp} showed that FSDP communication can be captured this way: storing parameters as sharded DTensors and triggering all-gather via parametrization makes the collectives ordinary graph nodes visible to the compiler; we generalize this from a fixed sharding rule to per-operator placement search (\S\ref{sec:strategy-space}). \textbf{TorchInductor}~\citep{ansel2024pytorch} compiles the resulting graph into fused kernels and schedules communication-computation overlap. Because AutoParallel produces an ordinary PyTorch module with DTensor parameters, the output flows through this stack unchanged: Inductor compilation, distributed checkpointing, mixed precision, and activation checkpointing all compose without adaptation. \textbf{GraphTrainer}~\citep{graphtrainer} is TorchTitan's full-graph training backend. AutoParallel uses it as backend and an entry point to standard Inductor compilation.

\subsection{\texorpdfstring{Challenges}{Challenges}}\label{sec:motivation}

The per-operator placement view (\S\ref{sec:placement-view}) makes the plan space expressive, but it raises three challenges for automatic parallelization in production PyTorch training.

\textbf{C1: Placement decisions are fine-grained and workload-specific.} Even with the mesh and parallelism degrees fixed, operator placements alone matter: constraining one or two projections of LLaMA~3 8B on a $4{\times}8$ mesh and re-solving yields 11 feasible plans whose measured step times span 2.2$\times$, from 365 to 790\,ms (Appendix~B). The best plan also shifts with the workload; at 32K tokens, a fresh plan runs 51\% faster than one tuned for 2K-token sequences (\S\ref{sec:replanning}). A planner must therefore search per-operator placements over the whole training graph and be cheap enough to rerun.

\textbf{C2: The search space grows exponentially with mesh dimensionality.} Production training stacks HSDP, tensor, context, and expert parallelism into high-dimensional meshes whose axes differ in bandwidth by an order of magnitude. Because a placement is a $D$-tuple (\S\ref{sec:placement-view}), the candidate strategies per operator grow exponentially with the mesh dimensionality~$D$. For the LLaMA~3 8B model, options grow from 12.9K decision variables on a 1D mesh to 12.2M on 3D, which takes 30 minutes to solve, and to 523M on 4D, where the problem cannot be built within an hour (\S\ref{sec:planner-scalability}). High-dimensional meshes therefore call for approximate search.

\textbf{C3: Plans should compose with the production stack.} A plan is most useful in production when it runs on the same compiler stack as hand-written recipes, composes with activation checkpointing, mixed precision, and distributed checkpointing, and leaves specialized regions, such as custom context-parallel kernels and data-dependent mixture-of-experts routing, under user control. Engineers also need to inspect, constrain, and replay plans; our own evaluation relies on all three (\S\ref{sec:case-study}, \S\ref{sec:replanning}, Appendix~B). Systems with a separate runtime make these requirements hard to meet in PyTorch.

We address C1 with graph-wide placement search (\S\ref{sec:strategy-space}), C2 with a seeded MRF solver (\S\ref{sec:approximate-solver}), and C3 with native DTensor output and \texttt{local\_map} boundaries (\S\ref{sec:production-integration}).

\FloatBarrier

\section{System Overview}\label{sec:system-overview}

AutoParallel takes a single-GPU model definition and a device mesh as input and produces a parallelized PyTorch module as output. Two principles guide its design. First, every component builds on the PyTorch compilation and distribution stack described in \S\ref{sec:pytorch-stack}, eliminating the need for a separate intermediate representation or custom runtime. Second, we draw an explicit boundary between computation optimized automatically and computation the user controls directly, composing the two through placement contracts rather than attempting to absorb all distributed patterns into the search. Both principles address C3 in \S\ref{sec:motivation}.

\subsection{End-to-End Workflow}\label{sec:workflow}

Figure~\ref{fig:overview} summarizes the workflow. Given a model, a device mesh, and representative inputs, AutoParallel \textcircled{\scriptsize 1} traces the model once with TorchDynamo and AOTAutograd into a joint forward-backward ATen graph, which requires no GPU. It then \textcircled{\scriptsize 2} enumerates candidate DTensor strategies for every operator, each defining how the operator executes within a single SPMD program; because candidates come from DTensor's propagation rules, every plan the solver can select is semantically valid by construction (\S\ref{sec:strategy-space}). Next, \textcircled{\scriptsize 3} the solver selects one strategy per operator to minimize estimated computation and communication cost under placement constraints (\S\ref{sec:cost-model}, \S\ref{sec:approximate-solver}). Finally, \textcircled{\scriptsize 4} the plan is materialized as DTensor parameters plus explicit redistribution collectives, producing an ordinary PyTorch module that runs eagerly or through the PyTorch compiler stack (\S\ref{sec:production-integration}).

\subsection{Optimization Boundary}\label{sec:optimization-boundary}

Standard tensor operators, including matrix multiplications, normalizations, attention projections, and element-wise functions, participate in placement search. Regions with dynamic or specialized distributed behavior, such as mixture-of-experts routing and context-parallel attention, remain user-written through \texttt{local\_map} placement contracts, which the solver treats as fixed constraints (\S\ref{sec:composition}).

\subsection{User Interface}\label{sec:user-interface}

At its simplest, AutoParallel takes a model, a mesh, representative inputs, and the desired output placements:

\begin{center}
\begin{minipage}{0.76\linewidth}
\begin{lstlisting}[style=apcode,basicstyle=\ttfamily\small\color{metafg}]
mesh = init_device_mesh("cuda", (4, 8))
with torch.device("meta"):
    model = Transformer(config)
par_model = auto_parallel(model, mesh, sample_inputs=(x,),
                          out_shardings=(Shard(0), Replicate()))
par_model.to_empty(device="cuda")
par_model.init_weights()
\end{lstlisting}
\end{minipage}
\end{center}

\noindent Input placements come from representative DTensors; ordinary tensors are replicated. After weight initialization, the module supports a standard training loop. Users can constrain input/output placements and parameter storage, configure the solver, inspect operator costs, and serialize plans for replay. For auto-planned operators, mesh names are descriptive labels and placements are chosen freely across axes.

\FloatBarrier

\section{Placement Planning}\label{sec:solver}

We formulate parallelization as a graph-wide placement optimization over the joint forward-backward graph. Restricting the search to SPMD placements keeps the space tractable while still expressing data parallelism, FSDP, tensor and sequence parallelism, and their combinations; pipeline parallelism, which partitions the graph into stages rather than sharding operators, is orthogonal and discussed in \S\ref{sec:related-work}.

\subsection{Strategy Space}\label{sec:strategy-space}

The search space is built from PyTorch DTensor's placement vocabulary. On a $D$-dimensional device mesh, the placement of a tensor is a $D$-tuple, one element per mesh dimension, drawn from $\{\text{Replicate},\, \text{Shard}(d),\, \text{Partial}\}$ (\S\ref{sec:placement-view}). A \emph{strategy} for an operator specifies the placement of every input and output tensor; the set of valid strategies for an operator is determined by DTensor's propagation rules, which encode the semantics of each ATen operation. Each rule admits only input and output placements under which the per-device local computations, combined with the implied collectives, reproduce the result of the unsharded operator. Every candidate plan is therefore semantically equivalent to single-device execution up to floating-point reassociation in reductions, the same property that holds for hand-written data- and tensor-parallel recipes. Every enumerated strategy is valid in this sense, and so is any combination of them: plan application inserts the required collective wherever adjacent placements disagree (\S\ref{sec:materialization}), so the selected plan computes correct results and the planner can optimize purely for performance. Training with AutoParallel plans reproduces the expert recipe's loss curve within the variation caused by changing the mesh alone (Appendix~D).

\begin{table}[htbp]
\centering
\caption{Strategies for $Y\!=\!XW$ on a two-dimensional mesh. Each cell is a placement tuple with one entry per mesh axis. S($d$)\,=\,Shard($d$), R\,=\,Replicate, P\,=\,Partial, AG\,=\,AllGather, AR\,=\,AllReduce; (0): on mesh axis~0 only.}
\label{tab:strategy-example}
\normalsize
\begin{tabular*}{0.90\linewidth}{@{\extracolsep{\fill}}lcccc@{}}
\toprule
& $X$ & $W$ & $Y$ & Comm. \\
\midrule
\textbf{DP} & S(0),\,S(0) & R,\,R & S(0),\,S(0) & AR $\nabla\!W$ \\
\textbf{FSDP} & S(0),\,S(0) & S(0),\,S(0) & S(0),\,S(0) & AG $W$ \\
\textbf{Col-TP} & R,\,R & R,\,S(1) & R,\,S(1) & --- \\
\textbf{Row-TP} & R,\,S(1) & R,\,S(0) & R,\,P & AR $Y$ \\
\textbf{FSDP+TP} & S(0),\,R & S(0),\,S(1) & S(0),\,S(1) & AG $W$ (0) \\
\bottomrule
\end{tabular*}
\end{table}

To ground this abstraction, consider a matrix multiplication $Y = XW$ on a two-dimensional mesh, illustrated in Table~\ref{tab:strategy-example}. Sharding $X$ along the batch dimension across the entire mesh while replicating $W$ produces data parallelism; additionally sharding $W$ across the mesh produces FSDP (the table shows $W$ in its stored placement; FSDP all-gathers it before the multiplication). These patterns are not declared; they are the operational consequences of particular placement tuples. Combining FSDP on one axis with column-parallel TP on another produces the FSDP$\,+\,$TP strategy: the planner assigns each axis its role based on cost, not naming convention. The optimizer selects among all valid candidates based on their estimated cost on the target hardware.

In general, the candidate set for each operator is the Cartesian product of per-dimension placements, filtered by operator semantics. On a $D$-dimensional mesh, a tensor with $n$ dimensions admits up to $(1 + n)^D$ candidate placements (Replicate plus $\text{Shard}$ along each of $n$ dimensions, raised to the $D$th power across mesh dimensions). After semantic filtering, which removes placements that shard a dimension smaller than the mesh degree or that violate operator constraints, the typical count per node is 10--30 on a two-dimensional mesh. On meshes with three or more dimensions, the combinatorial growth makes even enumeration expensive, motivating the seeding technique described in \S\ref{sec:approximate-solver}. Operators wrapped in \texttt{local\_map} (\S\ref{sec:composition}) receive a single-strategy candidate set determined by the user's declared boundary placements, removing them from the search.

Given the per-node candidate sets, the placement problem is to select exactly one strategy $s_v \in S_v$ per node $v$ to minimize the sum of per-operator computation cost and per-edge redistribution cost:
\begin{equation}\label{eq:objective}
\min_{\{s_v\}} \;\sum_{v \in V} c_{\mathrm{comp}}(v,\, s_v) \;+\; \sum_{(u,v) \in E} c_{\mathrm{redist}}(s_u,\, s_v)
\end{equation}
\noindent where $V$ is the set of nodes and $E$ the set of edges in the computation graph, $c_{\mathrm{comp}}(v, s_v)$ is the computation cost of executing node~$v$ under strategy~$s_v$, and $c_{\mathrm{redist}}(s_u, s_v)$ is the communication cost of the redistribution collective required when the output placement of~$u$ disagrees with the input placement expected by~$v$ (zero when they agree). The optimization is subject to: (i)~uniqueness: each node selects exactly one strategy; (ii)~boundary constraints: placements of model inputs, outputs, and \texttt{local\_map} boundaries are fixed; (iii)~parameter-storage budget; and (iv)~forward-backward consistency: the forward and backward graphs must agree on the placements of tensors that cross the boundary between them, specifically model inputs and outputs and their corresponding gradients; intermediate activations within the forward or backward graph are not individually constrained. Both cost terms are estimated by the hardware-aware model described next.

\subsection{Hardware-Aware Cost Model}\label{sec:cost-model}

The cost model only needs to rank strategies in the same order as actual execution, so predicting relative costs matters more than predicting absolute time (evaluated in Appendix~B). It is also pluggable: users can recalibrate or replace it without changing the solver.

\textbf{Computation.} Each strategy implies a per-device tensor shape, to which the model applies a roofline estimate~\citep{williams2009roofline}: the larger of FLOPs over peak throughput and memory traffic over memory bandwidth, plus a minimum kernel-launch overhead so that many small operators are not underestimated.

\textbf{Communication.} Each redistribution maps to a collective: $\text{Shard}\to\text{Replicate}$ to all-gather, $\text{Partial}\to\text{Shard}$ to reduce-scatter, $\text{Partial}\to\text{Replicate}$ to all-reduce, and $\text{Shard}\to\text{Shard}$ across dimensions to all-to-all; $\text{Replicate}\to\text{Partial}$ is disallowed. A plain latency-bandwidth model misranks these collectives, because NCCL selects among algorithms (ring, tree, NVLS) and protocols by message size and topology, which can change effective bandwidth by an order of magnitude. To stay close to real execution, the model mirrors NCCL's selection logic, adds corrections calibrated against NCCL benchmarks on the target cluster, and uses per-axis bandwidth and latency, so that a large collective costs more on a slow inter-node axis than on a fast intra-node axis.

\subsection{Scalable Solver}\label{sec:approximate-solver}

The placement problem (Equation~\ref{eq:objective}) is an integer program whose difficulty grows with mesh dimensionality~$D$: the per-node search space grows as $O((1{+}n)^D)$ (\S\ref{sec:strategy-space}), making strategy enumeration and pairwise cost computation expensive even before the solver runs, and the combinatorial optimization over the full graph compounds this cost. As \S\ref{sec:motivation} shows, an exact ILP formulation with CBC~\citep{forrest2005cbc} becomes impractical beyond two dimensions; we keep it as a reference for solution quality (\S\ref{sec:planner-scalability}).

\textbf{From ILP to MRF.} The ILP formulation constrains parameter storage through the sum of the parameters' local-to-global size ratios. Parameter placement choices couple this budget across the graph. In production model training, however, parameter memory redundancy across devices is not acceptable: the standard practice is to fully shard all parameters across the non-DDP mesh dimensions, and this is the tightest packing any feasible plan can achieve. When we enforce parameter sharding in the candidate strategies, restricting weight nodes to $\text{Shard}$ placements on all non-DDP mesh axes, persistent parameter storage is fixed: each device stores exactly its $1/N$ share of every parameter, where $N$ is the product of non-DDP mesh degrees. The parameter-storage constraint then no longer depends on the remaining decisions. Appendix~C examines placement choices under different parameter-storage budgets. The ILP comparison in \S\ref{sec:planner-scalability} uses the same parameter-sharding restriction, so both solvers optimize the same objective. With inputs sharded and redundant computation charged by the cost model, the selected plans in our experiments keep activations sharded, and peak activation memory is governed by selective activation checkpointing.

With parameter storage fixed, the objective (Equation~\ref{eq:objective}) decomposes into per-node terms (computation cost) and per-edge terms (redistribution cost), each depending on at most two adjacent nodes' strategy choices. This structure is a pairwise Markov Random Field (MRF)~\citep{kolmogorov2006convergent}, enabling efficient approximate inference via message passing rather than integer programming.

\textbf{Making the MRF tractable.} Two reductions shrink the problem before solving.

\emph{Graph clustering.} Models with repeated structure, such as the $L$ identical layers of a transformer, produce $L$ copies of the same subproblem in the computation graph. We identify copies via content-based hashing of each node's operator, arguments, and strategy set, and share decision variables across identical groups.

\emph{Split-dimension seeding.} Even after clustering, candidate enumeration and pairwise cost computation grow exponentially with $D$ (Table~\ref{tab:solver-scalability}). We therefore decompose the $D$-dimensional problem into $D$ independent one-dimensional subproblems, one per mesh axis, each using a projected cost model reflecting that axis's bandwidth. Each 1D problem is small enough to solve exactly. The per-dimension solutions are assembled into a $D$-tuple \emph{seed} per node, and the full problem is rebuilt with enumeration restricted to a Hamming ball of radius~$r$ (default $r{=}2$) around each seed, retaining only strategies that differ from the seed in at most $r$ dimensions. This reduces the effective candidate count from $O((1{+}n)^D)$ to $O(D^r \cdot n^r)$, making enumeration and cost computation tractable at high~$D$.

\textbf{Solving the MRF.} Nodes that must choose jointly, including clustered copies, forward-backward pairs, and flow-coupled producers, are merged into groups via union-find, each group selecting one joint assignment. Flow-coupled producers feed the same input position of clustered operator copies and therefore share one placement decision. The unary potential for a group aggregates computation and self-redistribution costs across its members, weighted by the number of cluster copies each represents. The pairwise potential between adjacent groups is the redistribution cost between their selected strategies.

We solve the resulting MRF with Tree-Reweighted Sequential message passing (TRW-S), which alternates forward and backward sweeps over monotonic chains induced by a topological ordering and keeps the best decoded assignment across sweeps. Two rounds of local search then refine the solution: coordinate descent over single groups, and star-block search over high-degree groups together with their neighbors.

\FloatBarrier

\section{Execution and Composition}\label{sec:production-integration}

\subsection{Plan Materialization}\label{sec:materialization}

We materialize the placement plan through a two-pass lowering of the joint forward-backward graph. In the first pass, an interpreter walks every node and inserts redistribution collectives (all-gather, reduce-scatter, all-to-all) wherever the output placement of one operator disagrees with the input placement expected by the next. View and factory operations have their shape arguments adjusted from global to per-device local dimensions. In the second pass, standard Inductor decompositions lower any remaining compound operations into base ATen operators. Model parameters are distributed according to the plan: each parameter is wrapped as a DTensor with the selected placement, sharded across the device mesh. The lowered graph and sharded parameters are packaged into an ordinary \texttt{nn.Module} that inherits the original model's class, preserving \texttt{isinstance} checks and module hierarchy. The output is a standard PyTorch module with no custom runtime state, ready for training loops, eager debugging, or compilation.

\subsection{Compilation and Communication Overlap}\label{sec:compilation}

For production training, the complete forward, loss, and backward computation of the materialized module, including the redistribution collectives inserted during materialization, is traced into a single FX graph and compiled with Inductor through TorchTitan's GraphTrainer~\citep{graphtrainer}. We rely on generic communication reordering, overlap scheduling, and collective bucketing passes already present in Inductor rather than reimplementing scheduling logic in the planner.

\subsection{Composing with Dynamic Parallelism}\label{sec:composition}

Not all distributed computation can or should be folded into static placement search: some patterns involve dynamic collectives, and others benefit from user-controlled implementations that have been tuned for specific hardware or workloads. We handle both cases through \texttt{local\_map}, a PyTorch higher-order operator that defines a placement contract at the composition boundary: the user declares how inputs should be distributed before entering the region and how outputs are distributed when exiting, then implements the internal computation and collectives using standard PyTorch distributed primitives. The optimizer treats these declared placements as fixed single-strategy constraints, optimizing all surrounding computation to be compatible with the boundary.

\begin{figure}[htbp]
\centering
\includegraphics[width=\linewidth]{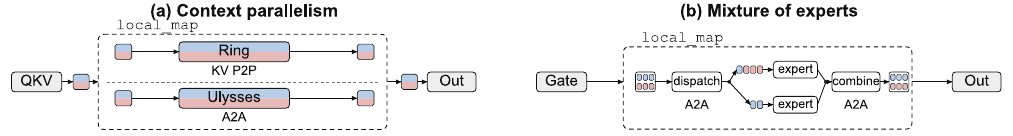}%
\caption{Composition through \texttt{local\_map}. Dashed boxes are user-written \texttt{local\_map} regions; AutoParallel plans the operators outside them (gray).}
\label{fig:composition-new}
\end{figure}

\textbf{Context parallelism.} Context parallelism illustrates a region kept user-written because it relies on a custom implementation. For long-context training, context parallelism distributes the sequence dimension across devices so that each device processes a chunk of the sequence. We wrap the attention computation in \texttt{local\_map} with input placements that shard query, key, and value tensors along the sequence dimension on the CP mesh axis, allowing users to plug in their preferred implementation (Figure~\ref{fig:composition-new}a).

\textbf{Mixture-of-experts.} Mixture-of-experts illustrates the other reason: its routing produces data-dependent shapes that static placement search cannot express. Expert parallelism distributes experts across devices with data-dependent token routing and all-to-all dispatch. We construct a device mesh where the expert-parallel axis is folded into the existing mesh dimensions and wrap the token-dispatch, expert-computation, and token-combine phases in \texttt{local\_map} with placements declaring which axis carries expert parallelism (Figure~\ref{fig:composition-new}b).

In both cases the composition boundary is explicit and inspectable: the user controls the internal implementation and placement declarations, and the optimizer handles everything outside.

\FloatBarrier

\section{Evaluation}\label{sec:evaluation}

We compare AutoParallel with expert-written recipes (\S\ref{sec:end-to-end-performance}), examine the placement decisions behind the differences (\S\ref{sec:case-study}), measure the benefit of replanning when the workload changes (\S\ref{sec:replanning}), and evaluate solver scalability on high-dimensional meshes (\S\ref{sec:planner-scalability}). Appendix~B evaluates the cost model's ranking fidelity.

\FloatBarrier

\subsection{Setup}\label{sec:setup}

All experiments run on nodes each with eight NVIDIA H100 80\,GB SXM GPUs connected via NVLink within a node and RoCE across nodes. The software stack is PyTorch 2.14.0a0, CUDA 13.0, NCCL 2.30.7, and Python 3.12. All configurations train in \texttt{bfloat16} mixed precision with \texttt{float32} reductions; \texttt{torch.compile}, selective activation checkpointing policy, and communication settings are matched across arms at every mesh.

We evaluate LLaMA~3 8B and 70B~\citep{llamateam2024llama3}, Qwen3 32B~\citep{yang2025qwen3}, and MuseGlimmer 30B~\citep{metamuseglimmer}, whose hybrid attention combines sliding-window and full attention. Our 16B MoE configuration uses the DeepSeek-V3 architecture~\citep{deepseek2024v3} with fewer layers and heads.

The baseline is \textbf{TorchTitan MainTrainer}~\citep{liang2024torchtitan}: expert-designed parallel strategies executed with the standard PyTorch training loop. \textbf{AutoParallel} receives the logical mesh and boundary placement contracts for \texttt{local\_map} regions (context parallelism, mixture-of-experts) but no operator-level placement hints; it discovers the parallel strategy automatically and executes through GraphTrainer with full-graph Inductor compilation, communication-computation overlap, and collective bucketing. Both arms use the full set of Inductor optimizations: MainTrainer through JIT \texttt{torch.compile} and AutoParallel through a single full-graph trace. In GraphTrainer, AutoParallel keeps the default graph cleanup and selective activation checkpointing passes but disables GraphTrainer's own collective bucketing, overlap-scheduling; communication is instead scheduled by Inductor's overlap scheduling and collective bucketing. We measure a complete training iteration, excluding data loading and one-time tracing, planning, and compilation. We run 5 warm-up iterations followed by 20 measured iterations and report average throughput in tokens per second.

\FloatBarrier

\subsection{End-to-End Performance}\label{sec:end-to-end-performance}

\begin{table}[htbp]
\centering
\caption{End-to-end throughput (K tokens/s) on scaling ladders from 8 to 128 GPUs. $\Delta$ is AutoParallel relative to MainTrainer. All runs use local batch size 2 per data-parallel rank.}
\label{tab:e2e-results}
\normalsize
\begin{tabular*}{0.95\linewidth}{@{\extracolsep{\fill}}l r rr r@{}}
\toprule
Mesh & GPUs & MT & AP & $\Delta$\% \\
\midrule
\multicolumn{5}{@{}l}{\textit{LLaMA~3 8B (seq 8\,K; 4D uses 16\,K with CP)}} \\
$2{\times}4$ & 8 & 54.2 & 58.4 & +7.9 \\
$2{\times}8$ & 16 & 82.1 & 88.0 & +7.2 \\
$4{\times}8$ & 32 & 165.3 & 181.5 & +9.8 \\
$2{\times}4{\times}8$ & 64 & 323.8 & 335.4 & +3.6 \\
$4{\times}4{\times}2{\times}4$ & 128 & 575.5 & 554.7 & $-$3.6 \\
\midrule
\multicolumn{5}{@{}l}{\textit{DeepSeek-V3 16B MoE (seq 8\,K)}} \\
$1{\times}8$, EP\,8 & 8 & 44.2 & 52.3 & +18.3 \\
$4{\times}8$, EP\,8 & 32 & 153.5 & 162.6 & +5.9 \\
$8{\times}8$, EP\,8 & 64 & 280.6 & 285.1 & +1.6 \\
$4{\times}4{\times}8$, EP\,32 & 128 & 349.8 & 373.5 & +6.8 \\
\midrule
\multicolumn{5}{@{}l}{\textit{MuseGlimmer 30B Hybrid (seq 4\,K)}} \\
$8{\times}2$ & 16 & 36.0 & 36.2 & +0.5 \\
$16{\times}2$ & 32 & 72.3 & 72.0 & $-$0.4 \\
$32{\times}2$ & 64 & 142.1 & 137.3 & $-$3.4 \\
\midrule
\multicolumn{5}{@{}l}{\textit{Qwen3 32B Dense (seq 8\,K)}} \\
$16{\times}8$ & 128 & 199.7 & 195.1 & $-$2.3 \\
\midrule
\multicolumn{5}{@{}l}{\textit{LLaMA~3 70B Dense (seq 8\,K)}} \\
$16{\times}8$ & 128 & 101.0 & 99.2 & $-$1.8 \\
\bottomrule
\end{tabular*}
\end{table}

Table~\ref{tab:e2e-results} reports aggregate throughput on scaling ladders from 8 to 128 GPUs. Each ladder increases mesh dimensionality with GPU count, from two-dimensional dp$\times$tp at small scale to three- and four-dimensional meshes at larger scales. Mesh shapes list axes outermost first: the LLaMA~3 8B 3D and 4D tuples are (DDP,\,DPS,\,TP) and (DDP,\,DPS,\,CP,\,TP), respectively. EP gives the expert-parallel degree folded into the mesh; in DeepSeek-V3, EP\,8 uses the TP axis, while EP\,32 spans the final two axes of the $4{\times}4{\times}8$ mesh. Across the 14 configurations, AutoParallel outperforms MainTrainer on 9 and stays within 3.6\% on the remaining 5.

\textbf{Graph-wide search discovers non-obvious placements.} On LLaMA~3 8B with two-dimensional dp$\times$tp meshes (8--32 GPUs), AutoParallel consistently outperforms the expert baseline, improving throughput by 7.2--9.8\%. The graph-wide search discovers per-operator placement refinements that the uniform manual strategy does not explore: AutoParallel tailors placements to individual operators within each layer based on their computation and communication costs on the target mesh. Because the search spans the joint forward-backward graph, the forward and backward computations of the same layer can also take different layouts, which module-level recipes do not express. \S\ref{sec:case-study} examines these plans in detail. On DeepSeek-V3 16B, where \texttt{local\_map} controls expert routing and AutoParallel optimizes the surrounding attention and shared-expert layers, the improvement reaches 18.3\% at 8 GPUs ($1{\times}8$, EP\,8). These gains likely come from per-operator placements in the attention and shared-expert layers surrounding the \texttt{local\_map} block, which the uniform expert recipe keeps fixed. On MuseGlimmer 30B, LLaMA~3 70B, and Qwen3 32B, AutoParallel matches expert throughput within 3.4\%, indicating that the expert recipes for these models are already close to the plans AutoParallel selects, which it reaches without per-model tuning. Notably, Qwen3 is an external model family with no AutoParallel-specific recipe engineering, demonstrating generalization beyond the models used during development.

\textbf{Plans scale to high-dimensional meshes.} As GPU count grows, production training moves to higher-dimensional meshes that combine HSDP, tensor, context, and expert parallelism. On LLaMA~3 8B at 64 GPUs with a three-dimensional HSDP+TP mesh ($2{\times}4{\times}8$), AutoParallel improves throughput by 3.6\% over the expert baseline. At 128 GPUs with a four-dimensional HSDP+CP+TP mesh ($4{\times}4{\times}2{\times}4$), which adds context parallelism via \texttt{local\_map} for 16K sequences, AutoParallel is within 3.6\% of the expert. Context parallelism places most of the attention computation inside the manually controlled region, limiting the scope of auto-optimization to the surrounding projections and feed-forward layers; we attribute the bounded gap mainly to this smaller optimization scope, a trade-off we accept by design, although we have not isolated it from other factors. The same pattern holds on DeepSeek-V3, where the mesh transitions from two-dimensional dp$\times$ep at 8--64 GPUs to three-dimensional ($4{\times}4{\times}8$, EP\,32) at 128 GPUs, and AutoParallel achieves a 6.8\% improvement at the highest dimensionality.

\FloatBarrier

\subsection{\texorpdfstring{Discovered Plans}{Discovered Plans}}\label{sec:case-study}

We inspect two plans: one that reproduces the expert recipe and one whose departures from it are instructive. We highlight individual decisions rather than enumerate every difference. We write $W_{qkv}$ for the query, key, and value projections collectively, $W_o$ for the attention output projection, $W_1$/$W_3$ for the gate/up projections of the gated MLP, and $W_2$ for its output projection. SDPA denotes scaled dot-product attention.

\textbf{MuseGlimmer $\mathbf{8{\times}2}$: recovering the expert plan.} On every MuseGlimmer mesh, AutoParallel independently arrives at the same strategy as MainTrainer: FSDP on the dp axis, Megatron-style column/row tensor parallelism with sequence parallelism on tp, and attention heads sharded across tp. There are zero role-level deviations across all operators. This indicates that on this model the expert recipe is already close to the cost-model optimum, and the solver reaches it without guidance.

\begin{figure}[htbp]
\centering
\includegraphics[width=\linewidth]{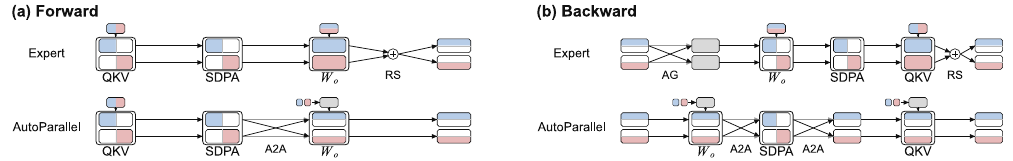}%
\caption{Schematic attention-block placement of the expert plan and AutoParallel. Vertical and horizontal splits denote head and sequence sharding.}
\label{fig:wo-new}
\end{figure}

\textbf{LLaMA~3 8B: mixing Megatron TP with Ulysses-style all-to-all.} Figure~\ref{fig:wo-new} shows the attention block of the plan AutoParallel selects on the 128-GPU $16{\times}8$ mesh. In the forward pass, instead of applying row-parallel $W_o$ and reduce-scattering the partial sum, AutoParallel converts the head-sharded attention output to sequence-sharded with an all-to-all, as in DeepSpeed-Ulysses~\citep{jacobs2023deepspeed}, and applies $W_o$ with the full, all-gathered weight. With $A$ the activation size of one TP group and $p$ the TP degree, the all-to-all sends about $A(p{-}1)/p^2$ bytes per rank, a factor of $p$ less than the reduce-scatter; the plan pays an extra weight gather, which Inductor can overlap with computation. The same $W_o$ change appears on the $4{\times}8$ and $2{\times}4{\times}8$ meshes, including the $4{\times}8$ configuration with the largest LLaMA gain (9.8\%), although we have not isolated it from the rest of the plan. Each ingredient (Megatron TP, the Ulysses head-to-sequence all-to-all, and FSDP-style weight gathering) is individually well known, but the evaluated uniform TorchTitan recipe does not mix them at a single operator within a single layer; the per-operator search finds this combination automatically.

\textbf{Different forward and backward layouts.} The backward paths in Figure~\ref{fig:wo-new} show a second departure. Module-level recipes assign each projection one TP role, which fixes both its forward layout and its backward communication; AutoParallel places forward and backward operators independently and constrains only tensors that cross the forward-backward boundary (\S\ref{sec:strategy-space}). It keeps the Q/K/V outputs head-sharded in the forward pass, as SDPA requires, but runs their grad-input matmuls sequence-sharded, so two all-to-alls replace the expert's backward all-gather and reduce-scatter. The cost model estimates a saving of about 0.34\,ms of backward communication per layer, which the current implementation partly loses to weight gathers over the combined mesh.

These two examples are representative, not exhaustive: other configurations exhibit further deviations of varying complexity, but the pattern is consistent. On MuseGlimmer, the solver recovers the expert strategy; on LLaMA, it combines operator placements that differ from the uniform recipe. We note that throughput differences also reflect execution-level factors (compilation, bucketing, communication-computation overlap) that are not isolated in these comparisons.

\FloatBarrier

\subsection{Adaptation Through Replanning}\label{sec:replanning}

A key motivation for automatic planning is that workload changes invalidate stale recipes. We test this by generating a reference plan~$P_0$ for LLaMA~3 8B on a $4{\times}8$ mesh at sequence length 2{,}048 and evaluating two strategies as the sequence length increases: \emph{Replay} reuses $P_0$'s placements unchanged, and \emph{Fresh} invokes AutoParallel with the new sequence length.

\begin{figure}[htbp]
\centering
\includegraphics[width=0.48\linewidth]{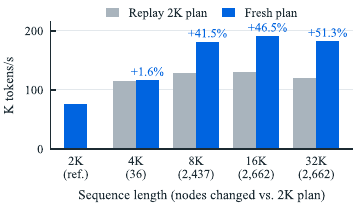}%
\caption{Throughput of fresh and replayed plans on LLaMA~3 8B ($4{\times}8$ mesh) across sequence lengths.}
\label{fig:replanning-line}
\end{figure}

Figure~\ref{fig:replanning-line} shows the results; labels give the throughput gain of Fresh over Replay. At 4K, only 36 nodes change placement and the gain is modest (1.6\%). As the sequence length grows further from the reference, the stale plan falls increasingly behind the fresh plan: at 8K, Fresh improves by 41.5\% with 2{,}437 changed nodes; at 32K, the advantage reaches 51.3\%. The widening gap reflects the coupled nature of parallelization decisions: as activation sizes grow, the memory-throughput balance shifts and placements optimized for 2K become inappropriate. Planning on this 2D mesh takes under 40\,s (\S\ref{sec:planner-scalability}), negligible relative to production runs of tens of thousands of steps.

\FloatBarrier

\subsection{Solver Scalability}\label{sec:planner-scalability}

We evaluate the approximate solver (\S\ref{sec:approximate-solver}) against the exact ILP on LLaMA~3 8B across one- through four-dimensional meshes, all with 32 or fewer GPUs so that the ILP remains verifiable where possible. We report build time (strategy enumeration and pairwise cost computation) and solve time separately. Graph clustering is enabled in both paths.

\begin{table}[htbp]
\centering
\caption{Solver scalability on LLaMA~3 8B. ILP vars: decision variables of the unseeded ILP; T/O: ILP build exceeds one hour; objectives in $\mu$s.}
\label{tab:solver-scalability}
\normalsize
\begin{tabular*}{\linewidth}{@{\extracolsep{\fill}}l r rr rr rr@{}}
\toprule
& & \multicolumn{2}{c}{Build (s)} & \multicolumn{2}{c}{Solve (s)} & \multicolumn{2}{c}{Objective} \\
\cmidrule(lr){3-4} \cmidrule(lr){5-6} \cmidrule(lr){7-8}
Mesh & ILP vars & ILP & Approx & ILP & Approx & ILP & Approx \\
\midrule
$1{\times}8$ & 12.9K & 4.7 & 4.9 & 0.3 & 5.1 & 626{,}547 & 626{,}547 \\
$4{\times}8$ & 341K & 17.9 & 12.4 & 12.3 & 6.1 & 311{,}290 & 311{,}290 \\
$4{\times}2{\times}4$ & 12.2M & 428.9 & 64.3 & 1{,}795 & 10.4 & 300{,}156 & 300{,}156 \\
$2{\times}2{\times}2{\times}4$ & 522.7M & T/O & 194.3 & --- & 20.8 & --- & 301{,}828 \\
\bottomrule
\end{tabular*}
\end{table}

Table~\ref{tab:solver-scalability} reports three results. \textbf{Quality.} On these 1D--3D instances, the approximate solver reaches the same objective as the ILP. \textbf{Scaling.} Build and solve are separate bottlenecks: seeding cuts the 3D build time by 6.7$\times$, and MRF inference keeps solve time under 21\,s on every mesh, so the approximate solver finishes the 3D mesh 30$\times$ faster than the ILP; on the 4D mesh the ILP cannot finish building within an hour. On the trivially small 1D mesh, the approximate solver's fixed overhead makes it slightly slower. \textbf{End-to-end cost.} Including one-time graph tracing, planning takes 25--233\,s from 1D to 4D, a one-time cost amortized over training runs that typically span hours to weeks.

\FloatBarrier

\section{Related Work}\label{sec:related-work}

\paragraph{Operator-level automatic parallelism.}
FlexFlow~\citep{jia2019beyond} searches a per-operator parallelism space with MCMC simulation over the Legion runtime; Unity~\citep{unger2022unity} extends this with joint algebraic-parallel optimization. Alpa~\citep{zheng2022alpa} formulates intra-operator sharding as an ILP and inter-operator pipeline partitioning as dynamic programming, targeting JAX/XLA with a Ray runtime. PartIR~\citep{alabed2025partir} decomposes parallelization strategies into composable tactics with optional MCTS-based search, using an MLIR-based partitioning stack; users can inspect the rewritten module after each tactic. On the propagation side, GSPMD~\citep{xu2021gspmd, lepikhin2020gshard} completes a full-graph sharding from sparse user annotations via heuristic propagation without a cost model; it supports manually partitioned subgraphs in XLA and has been deployed at production scale, but the key placement decisions are made by the user rather than a solver. These systems use compiler stacks and runtimes outside native PyTorch execution, and their partitioned programs do not directly compose with the PyTorch compilation ecosystem.

\paragraph{Layer-level planners.}
A second line of work plans at layer or stage granularity, choosing among named parallelism strategies such as DP, TP, PP, and ZeRO. Galvatron~\citep{miao2023galvatron} uses decision-tree pruning and dynamic programming to assign a strategy per layer. Mist~\citep{zhu2025mist} and UniAP~\citep{chen2024uniap} plan under memory constraints via MILP and MIQP, respectively. These systems operate at a granularity too coarse to express the per-operator mixed placements from which strategies like FSDP and TP emerge naturally in our formulation (\S\ref{sec:strategy-space}), and they rely on Megatron or custom runtimes. Many of them also search pipeline parallelism, which AutoParallel does not. PP targets regimes where inter-node bandwidth is insufficient to overlap FSDP and TP communication, whereas AutoParallel targets the SPMD placement decisions within a pipeline stage. The two problems are orthogonal and can be composed hierarchically, as in Alpa's two-level design~\citep{zheng2022alpa}.

\paragraph{PyTorch-based systems.}
In the PyTorch ecosystem, TorchTitan~\citep{liang2024torchtitan} and veScale~\citep{li2025vescale} build on DTensor but rely on hand-written parallelization plans. Colossal-Auto~\citep{liu2023colossal} adapts Alpa's ILP to torch.fx but analyzes only the forward graph, outputs to a custom sharded module rather than DTensor, and does not integrate with torch.compile. nnScaler~\citep{lin2024nnscaler} generates multi-device PyTorch source code, but extending its search space requires expert-written constraints on its primitives, and it depends on its own runtime, which adds integration work when composing with the PyTorch compilation stack. JAX's \texttt{shard\_map}~\citep{jaxshardmap} provides a similar boundary for manually partitioned computation. AutoParallel embeds \texttt{local\_map} regions as fixed-placement constraints inside the global solver (\S\ref{sec:composition}). Table~\ref{tab:related} summarizes the key distinctions. To our knowledge, AutoParallel is the first system that automatically plans whole-graph operator placements for unmodified PyTorch models and emits a native PyTorch program composable with the production compilation stack.

\begin{table}[htbp]
\centering\normalsize
\caption{Comparison with representative parallelization systems. \emph{Auto}: searches for a plan automatically. \emph{Op}: per-operator decisions. \emph{PT in}: takes an unmodified PyTorch \texttt{nn.Module} (\texttt{local\_map} regions optional). \emph{No RT}: no additional runtime. \emph{Stack}: composes with the compiler, activation checkpointing, mixed precision, and distributed checkpointing. \emph{Comp.}: composes with hand-written dynamic regions such as CP and MoE. $\dagger$\,XLA-based. $\ddagger$\,Expert-defined search space.}
\label{tab:related}
\begin{tabular*}{0.92\linewidth}{@{\extracolsep{\fill}}lcccccc@{}}
\toprule
 & Auto & Op & PT in & No RT & Stack & Comp. \\
\midrule
TorchTitan & \xmark & {--} & \cmark & \cmark & \cmark & \cmark \\
GSPMD\,$^\dagger$ & \xmark & \cmark & \xmark & \cmark & \cmark & \cmark \\
Alpa & \cmark & \cmark & \xmark & \xmark & \xmark & \xmark \\
Galvatron & \cmark & \xmark & \cmark & \xmark & \xmark & \xmark \\
nnScaler\,$^\ddagger$ & \cmark & \cmark & \cmark & \xmark & \xmark & \xmark \\
\textbf{AutoParallel} & \cmark & \cmark & \cmark & \cmark & \cmark & \cmark \\
\bottomrule
\end{tabular*}
\end{table}

\FloatBarrier

\section{Conclusion}\label{sec:conclusion}

We presented AutoParallel, a PyTorch-native automatic parallelism planner that jointly optimizes per-operator placements across the full training graph, produces ordinary PyTorch programs without a separate runtime, and composes with both the PyTorch compilation stack and user-controlled dynamic parallelism regions. On up to 128 H100 GPUs, AutoParallel matches or exceeds expert-written TorchTitan recipes on most configurations and discovers plans that uniform expert recipes do not express. Crafting efficient parallelization strategies for distributed training has historically required scarce systems expertise, and we believe AutoParallel represents a step toward making high-performance distributed training accessible to the broader PyTorch community.

\clearpage
\beginappendix

\FloatBarrier

\section{Placement Solutions on Community Models}\label{app:hf-solutions}

We apply AutoParallel to 17 Hugging Face models on a single-node $1{\times}8$ H100 mesh and compare the cost-model objective against SimpleFSDP~\cite{zhang2024simplefsdp} as a zero-configuration baseline. All numbers are cost-model objectives ($\mu$s, no overlap modeling), not measured step times; no GPU execution is performed.

\paragraph{Setup.} Global batch size is 16 (local 2 per rank); inputs are constrained to $\text{Shard}(0)$. Mixed precision uses \texttt{bfloat16} compute with \texttt{float32} reductions. Sequence length is 4096 by default, capped at each model's position-table limit where applicable. The parameter-storage constraint uses $\texttt{high} = 1/\text{world\_size}$ (0.125), yielding the minimum feasible storage ratio for each parameter. AutoParallel solves the ILP exactly with CBC (status Optimal on all 17 instances, solve time 0.3--1.2\,s). Both arms share the same cost model, strategy space, and parameter-storage budget; the only difference is the placement assignment.

Table~\ref{tab:hf-solutions} shows the results. AutoParallel reduces the cost-model objective on all 17 models (0.2--40.8\%). The plans fall into three groups.

\textbf{FSDP with embedding/head optimization} (8 models, 559M--7.1B). The repeated transformer layers use the same all-gather/reduce-scatter pattern as SimpleFSDP; AutoParallel only changes the embedding and LM-head layers, storing the token embedding sharded along the vocabulary dimension and, in 6 of 8 models, applying column-parallel sharding to the LM head. The gains are 0.2--1.1\%.

\textbf{Tensor parallelism} (4 models, 67M--355M). AutoParallel replaces per-parameter collectives with activation collectives throughout the model: column-parallel QKV and FFN projections, row-parallel output projections with all-reduce, replicated layer norms. Communication cost drops by 21--41\,K\,$\mu$s at the expense of a small compute increase. The large gains (22--41\%) reflect the per-collective latency floor on NVSwitch ({$\sim$}160\,$\mu$s per all-gather/reduce-scatter): SimpleFSDP issues 211--585 small per-parameter collectives whose cost is 80--84\% latency overhead.

\textbf{Mixed} (5 models, 61M--6.1B). AutoParallel applies different strategies to different operators within the same layer. For example, on GPT-J 6B, attention runs data-parallel while the MLP uses column/row tensor parallelism; on T5 Base, forward weights are gathered for data-parallel compute but nearly half of the weight gradients are produced sharded without a reduce-scatter. Gains range from 0.9\% to 9.3\%.

The cost model does not model overlap or collective bucketing; Inductor's bucketing would reduce SimpleFSDP's latency-floor overhead in practice. These results show that AutoParallel generalizes across a broad range of popular architectures, from 61M to 7.1B parameters, consistently finding plans at least as good as SimpleFSDP without any per-model tuning, and discovering tensor-parallel or mixed strategies that a uniform FSDP baseline cannot express.

\begin{table}[H]
\centering
\caption{Cost-model objective change (\%) for AutoParallel vs.\ SimpleFSDP on 17 Hugging Face models ($1{\times}8$ H100 mesh); negative is better.}
\label{tab:hf-solutions}
\normalsize
\begin{tabular*}{0.90\linewidth}{@{\extracolsep{\fill}}llrr@{}}
\toprule
Model & Params & Seq & $\Delta$\% \\
\midrule
\multicolumn{4}{@{}l}{\textit{FSDP with embedding/head optimization}} \\
BLOOM 560M      & 559M  & 4096 & $-$1.1 \\
BLOOM 7B        & 7.1B  & 4096 & $-$0.9 \\
StableLM 1.6B   & 1.6B  & 4096 & $-$0.9 \\
Phi-1           & 1.4B  & 4096 & $-$0.6 \\
GPT-Neo 2.7B    & 2.7B  & 2048 & $-$0.5 \\
Phi-2           & 2.8B  & 4096 & $-$0.2 \\
Falcon 7B       & 6.9B  & 4096 & $-$0.2 \\
Pythia 6.9B     & 6.9B  & 4096 & $-$0.2 \\
\midrule
\multicolumn{4}{@{}l}{\textit{Tensor parallelism}} \\
DistilBERT      & 67M   & 512  & $-$40.8 \\
RoBERTa         & 125M  & 512  & $-$40.4 \\
BERT            & 110M  & 512  & $-$40.2 \\
GPT-2 Medium    & 355M  & 1024 & $-$22.3 \\
\midrule
\multicolumn{4}{@{}l}{\textit{Mixed strategies}} \\
GPT-2 XL        & 1.6B  & 1024 & $-$9.3 \\
T5 Small        & 61M   & 4096 & $-$3.1 \\
GPT-J 6B        & 6.1B  & 2048 & $-$3.0 \\
Pythia 160M     & 162M  & 4096 & $-$1.0 \\
T5 Base         & 223M  & 4096 & $-$0.9 \\
\bottomrule
\end{tabular*}
\end{table}

\FloatBarrier

\section{Cost Model Ranking Fidelity}\label{app:cost-model-fidelity}

We validate the cost model's ranking quality on LLaMA~3 8B with a $4{\times}8$ mesh (FSDP\,4, TP\,8; 32 H100 GPUs, sequence length 8192, \texttt{bfloat16} mixed precision). We generate 11 feasible placement plans that share the same mesh but differ in per-operator placements across the 7{,}803-node computation graph. Starting from the ILP-optimal plan~(p00), computed with only input, output, and memory constraints, we produce alternatives by fixing the output placement of one or two key projections ($W_{qkv}$, $W_o$, or $W_2$) to a specific strategy and re-solving the full ILP; the solver re-optimizes all remaining nodes jointly. Table~\ref{tab:cost-model-fidelity} lists the resulting projection-output placements and cost-model objectives. All 11 candidates run on the same four-node allocation with the same checkpoint, data, \texttt{torch.compile} configuration, and selective activation checkpointing; only the placement assignment differs.

\begin{table}[H]
\centering
\caption{Placement plans and cost-model ranking on LLaMA~3 8B ($4{\times}8$ mesh, 32 GPUs). Each plan assigns placements to all 7{,}803 graph nodes; the table shows outputs of the Q/K/V ($W_{qkv}$), attention output ($W_o$), MLP gate/up ($W_{1/3}$), and MLP output ($W_2$) projections. Placement tuples are (dp,\,tp); S$_d$\,=\,Shard($d$), R\,=\,Replicate, P\,=\,Partial. $C$: cost-model objective (K\,$\mu$s); $T$: measured step time (ms); rk: rank.}
\label{tab:cost-model-fidelity}
\normalsize
\begin{tabular*}{\linewidth}{@{\extracolsep{\fill}}l cccc r@{\,}r r@{\,}r@{}}
\toprule
 & \multicolumn{4}{c}{Output placement (dp, tp)} & \multicolumn{2}{c}{Pred.} & \multicolumn{2}{c}{Meas.} \\
\cmidrule(lr){2-5} \cmidrule(lr){6-7} \cmidrule(lr){8-9}
 & $W_{qkv}$ & $W_o$ & $W_{1/3}$ & $W_2$ & $C$ & rk & $T$ & rk \\
\midrule
p00 & S$_0$,S$_2$ & S$_0$,S$_1$ & S$_0$,S$_2$ & S$_0$,P & 642 & 1 & 364.9 & 1 \\
p01 & S$_0$,S$_2$ & S$_0$,S$_1$ & S$_2$,S$_1$ & P,S$_1$ & 694 & 2 & 494.8 & 5 \\
p02 & S$_0$,S$_2$ & P,S$_1$ & S$_0$,S$_2$ & S$_0$,P & 703 & 3 & 421.9 & 2 \\
p03 & S$_0$,S$_2$ & S$_0$,S$_1$ & S$_2$,S$_1$ & P,S$_0$ & 707 & 4 & 507.0 & 6 \\
p04 & S$_0$,S$_2$ & P,S$_0$ & S$_0$,S$_2$ & S$_0$,P & 710 & 5 & 430.8 & 3 \\
p05 & S$_0$,S$_2$ & P,S$_2$ & S$_0$,S$_2$ & S$_0$,P & 722 & 6 & 453.3 & 4 \\
p06 & P,R & S$_0$,S$_1$ & S$_0$,S$_2$ & S$_0$,P & 773 & 7 & 600.5 & 7 \\
p07 & P,P & S$_1$,S$_2$ & S$_1$,S$_2$ & S$_1$,P & 835 & 8 & 638.7 & 9 \\
p08 & P,P & R,S$_1$ & S$_0$,S$_2$ & S$_0$,P & 860 & 9 & 790.4 & 11 \\
p09 & P,P & P,S$_0$ & S$_1$,S$_2$ & S$_1$,P & 868 & 10 & 618.7 & 8 \\
p10 & P,P & S$_2$,S$_2$ & S$_1$,S$_2$ & S$_1$,P & 921 & 11 & 784.0 & 10 \\
\bottomrule
\end{tabular*}
\end{table}

\begin{figure}[H]
\centering
\includegraphics[width=0.65\linewidth]{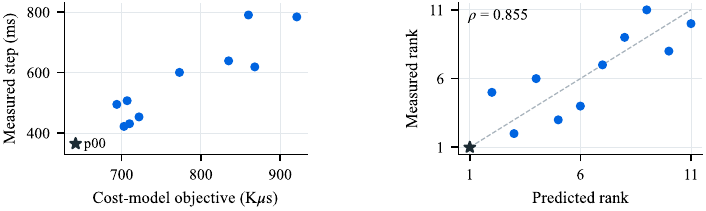}%
\caption{Left: Cost-model objective vs.\ measured step time for 11 placement plans on LLaMA~3 8B, $4{\times}8$ mesh; right: predicted vs.\ measured rank.}
\label{fig:cost-model-fidelity}
\end{figure}

Figure~\ref{fig:cost-model-fidelity} shows the results. The ILP-selected plan~(p00) is both the lowest-cost and the measured-fastest candidate (364.9\,ms), giving a selection regret of~0\%. Selection regret is $(T_{\mathrm{selected}}/\min_i T_i - 1)\times100\%$ over the 11 candidates. The Spearman rank correlation between cost-model objective and measured step time is $\rho = 0.855$ across the 11 candidates, which span a 2.2$\times$ range of measured times (365--790\,ms). The cost model does not model communication-computation overlap or collective bucketing; both are handled at runtime by Inductor and are likely contributors to the local ranking inversions visible in the middle of the range. The cost model's ranking quality is sufficient for plan selection among these candidates, even without modeling runtime optimizations.

\FloatBarrier

\section{Plan Stability Under Memory Budget}\label{app:memory-budget}

We sweep the normalized parameter-storage budget on LLaMA~3 70B with a $4{\times}8$ mesh (FSDP\,4, TP\,8; 32 GPUs, \texttt{bfloat16} mixed precision) from fully sharded ($\texttt{high} = 1/32$, i.e.\ each parameter stored at $1/\text{world\_size}$) to unconstrained ($\texttt{high} = 1$, every parameter allowed to be fully replicated). For this model, $\texttt{high}$ bounds the average local-to-global element-count ratio across parameter tensors. All results are ILP optimal; no GPU execution is performed. We additionally run three counterfactual solves at $\texttt{high} = 1$ that force all parameters to Replicate on the TP axis, the FSDP axis, or both.

\begin{table}[H]
\centering
\caption{Memory budget sweep on LLaMA~3 70B ($4{\times}8$ mesh). Placement tuples are (dp,\,tp).}
\label{tab:mem-sweep}
\normalsize
\begin{tabular*}{0.92\linewidth}{@{\extracolsep{\fill}}r rr cc r@{}}
\toprule
 & Obj.\ & Comm & \multicolumn{2}{c}{Norm placement} & Wt/GPU \\
\cmidrule(lr){4-5}
\texttt{high} & (K\,$\mu$s) & (K\,$\mu$s) & attn & ffn & (GB) \\
\midrule
1/32 & 3{,}854 & 1{,}878 & S$_0$,S$_0$ & S$_0$,S$_0$ & 8.819 \\
1/16 & 3{,}828 & 1{,}852 & R,S$_0$ & R,S$_0$ & 8.820 \\
1/8 & 3{,}826 & 1{,}850 & S$_0$,R & S$_0$,R & 8.820 \\
1/4 & 3{,}816 & 1{,}840 & S$_0$,R & R,R & 8.822 \\
1/2 & 3{,}807 & 1{,}831 & R,R & R,R & 8.824 \\
1 & 3{,}807 & 1{,}831 & R,R & R,R & 8.824 \\
\bottomrule
\end{tabular*}
\end{table}

Table~\ref{tab:mem-sweep} shows the results. Across the full budget range, every matrix projection ($W_{qkv}$, $W_o$, $W_1$/$W_3$, $W_2$, embedding, LM head) maintains identical placements, and compute cost is constant at 1{,}976\,K\,$\mu$s. The entire 1.2\% objective variation comes from the 161 norm-weight tensors (80 layers $\times$ 2 norms + final norm), which progressively transition from fully sharded to replicated as the budget increases. These 161 tensors account for 28.6\% of parameter tensor count but only 0.0019\% of parameter bytes (5.3\,MB of 282\,GB), so weights per GPU change by just 0.005\,GB.

\begin{table}[H]
\centering
\caption{Counterfactual solves forcing all parameters to Replicate on given mesh axes ($\texttt{high} = 1$). $\Delta$ is relative to the unconstrained baseline.}
\label{tab:mem-counterfactual}
\normalsize
\begin{tabular*}{0.82\linewidth}{@{\extracolsep{\fill}}l rr r@{}}
\toprule
Constraint & Obj.\ (K\,$\mu$s) & $\Delta$\% & Wt/GPU (GB) \\
\midrule
None & 3{,}807 & --- & 8.8 \\
R on tp & 3{,}911 & +2.7 & 70.6 \\
R on fsdp & 4{,}053 & +6.5 & 35.3 \\
R on both & 4{,}856 & +27.6 & 282.2 \\
\bottomrule
\end{tabular*}
\end{table}

Table~\ref{tab:mem-counterfactual} confirms why the solver does not replicate large matrices even when unconstrained. Forcing all parameters to Replicate on the TP axis, the FSDP axis, or both increases the objective by 2.7\%, 6.5\%, and 27.6\% respectively. The mechanism is gradient synchronization cost: sharded storage requires a \texttt{bfloat16} all-gather in the forward pass and a \texttt{float32} reduce-scatter for gradients, while replicated storage eliminates the forward gather but requires a \texttt{float32} all-reduce whose data volume is roughly double that of a reduce-scatter. At 70B-scale weight dimensions, this cost asymmetry dominates.

Norm weights behave differently because they have only 8{,}192 elements, making their communication cost latency-dominated. Sharded storage requires two cross-axis collectives per norm per layer (forward all-gather + backward reduce-scatter); replicated storage requires only one (backward all-reduce), saving approximately 294\,$\mu$s per norm. The total norm savings of 47{,}338\,$\mu$s account for the entire communication reduction from $\texttt{high} = 1/32$ to~1. In summary, the solver's placement decisions for the dominant parameters are driven by communication cost, not memory pressure, and are robust to the memory budget setting.

\FloatBarrier

\section{Numerical Validation}\label{app:numerics}

DTensor propagation rules restrict the search to plans that are semantically equivalent to single-device execution up to floating-point reassociation in reductions (\S4.1). To validate the materialized programs, we train LLaMA~3 8B for 1{,}000 steps on a DPS4$\times$CP2$\times$TP4 mesh (32 GPUs, sequence length 16K, global batch size 8, AdamW with learning rate $3\times10^{-4}$, 200 warm-up steps, and linear decay) with MainTrainer and with AutoParallel, starting from the same seed checkpoint and replaying the same data and learning-rate schedule. As a reference for how much reduction order alone changes the loss, we also run MainTrainer on the equivalent DPS4$\times$CP1$\times$TP8 mesh.

\begin{figure}[H]
\centering
\includegraphics[width=0.62\linewidth]{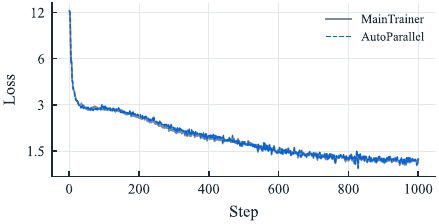}
\caption{Training loss of LLaMA~3 8B with MainTrainer and AutoParallel.}
\label{fig:loss}
\end{figure}

Figure~\ref{fig:loss} shows that the two loss curves overlap. Over the last 100 steps, AutoParallel's loss differs from MainTrainer's by 0.80\% on average, the same as MainTrainer on the TP8 mesh (0.82\%); both comparisons show larger transient deviations between steps 100 and 500. AutoParallel's numerical behavior is therefore within the variation that any change of parallel layout introduces.

\clearpage
\bibliographystyle{assets/plainnat}
\bibliography{paper}

\section*{AI Usage Disclosure}

We used AI assistants based on Claude to help with writing and language polishing, to launch and monitor experiment jobs, and to organize experimental data into tables and figures. All ideas, the system design and implementation, the experimental design, the structure and writing approach of the paper, and the analysis and interpretation of the data are our own. We reviewed all AI-assisted code and text, and we take full responsibility for the content of this paper.

\end{document}